\documentclass[aps,twocolumn,prb,showpacs,superscriptaddress]{revtex4}
\usepackage{amssymb}
\usepackage{epsfig}
\usepackage{epsf}

\def\frac#1#2{{\textstyle{#1 \over #2}}}
\def\half{\frac{1}{2}}

\def\be{\begin{equation}} \def\ee{\end{equation}}
\def\bea{\begin{eqnarray}} \def\eea{\end{eqnarray}}

\def\nn{\nonumber}

\def\ssr{\scriptscriptstyle\rm}
\def\ocite#1{[\onlinecite{#1}]}
\def\yd{^\dagger}
\def\nd{^{\vphantom{\dagger}}}
\def\xhat{{\hat x}}
\def\yhat{{\hat y}}
\def\zhat{{\hat z}}
\def\dnxy{d^{xy}_{s,a}}
\def\dnxz{d^{xz}_{s,a}}
\def\dnyz{d^{yz}_{s,a}}
\def\ddxy{d^{xy\,\dagger}_{s,a}}
\def\ddxz{d^{xz\,\dagger}_{s,a}}
\def\ddyz{d^{yz\,\dagger}_{s,a}}
\def\rvec{{\vec r}}
\def\rvecp{{\vec r}^{\,\prime}}
\def\kvec{{\vec k}}

\def\Qvec{{\vec Q}}

\begin{document}

\title{Quasiparticle Interference in the Unconventional Metamagnetic Compound Sr$_3$Ru$_2$O$_7$}
\author{Wei-Cheng Lee}
\email{leewc@physics.ucsd.edu}
\author{D. P. Arovas}
\email{darovas@ucsd.edu}
\author{Congjun Wu}
\email{wucj@physics.ucsd.edu}
\affiliation{Department of Physics, University of California, San Diego, California, 92093, USA}
\date{\today}

\begin{abstract}
Quasiparticle interference (QPI) in spectroscopic imaging scanning 
tunneling microscopy provides a powerful method to detect orbital band
structures and orbital ordering patterns in transition metal oxides.
We use the $T$-matrix formalism to calculate the QPI spectra for the 
unconventional metamagnetic system of Sr$_3$Ru$_2$O$_7$ with a 
$t_{2g}$-orbital band structure.
A detailed tight-binding model is constructed accounting for features such as
spin-orbit coupling, bilayer splitting, and the staggered rotation of the RuO octahedra.
The band parameters are chosen by fitting the calculated Fermi surfaces with 
those measured in the angular-resolved photo-emission spectroscopy experiment.
The calculated quasiparticle interference at zero magnetic field exhibits 
a hollow square-like feature arising from the nesting of the quasi-1d
$d_{xz}$ and $d_{yz}$ orbital bands, in agreement with recent  measurements
by J. Lee {\it et al.} (Nature Physics {\bf 5}, 800 (2009)).
Rotational symmetry breaking in the nematic metamagnetic state also 
manifests in the quasi-particle interference spectra.
\end{abstract}
\maketitle

\section {Introduction}
The physics of transition metal oxides is characterized by a rich
interplay among the lattice, charge, 
spin and orbital  degrees of freedom\cite{imada1998,tokura2000,khaliullin2005,khomskii2005}.
Various exotic phenomena, such as metal-insulator transitions and colossal magnetoresistance occur in orbitally active 
compounds with partially filled $d$ or $f$-shells.  In the literature many Mott-insulating orbital systems 
({\it e.g.\/},  La$_{1-x}$Sr$_x$MnO$_3$, La$_4$Ru$_2$O$_{10}$, LaTiO$_3$, 
YTiO$_3$, KCuF$_3$) \cite{murakami1998,ulrich2008,ichikawa2000,khalifah2002}
have been extensively studied, and both orbital ordering and orbital excitations have been observed.
Significant developments in orbital physics have also been made recently in cold atom optical lattice systems.
In particular, strongly correlated $p$-orbital bands filled with
both bosons and fermions provides a new perspective on orbital physics
which has not yet been explored in the solid state context
\cite{muller2007,liu2006,wu2006,wu2008,wu2008a,wu2009,lee2009a}.
In contrast, most $p$-orbital solid state systems exhibit only relatively weak correlations.
 
Metallic orbital systems, such as strontium ruthenates and 
iron-pnictide superconductors, have received a great deal of attention of late.
Their Fermi surfaces are characterized by hybridized $t_{2g}$-orbital bands, {\it i.e.}, the eigen-orbital 
admixture of the Bloch state varies around a connected region of the Fermi surface.
Orbital ordering in such systems corresponds to a preferred occupation along particular
directions on the Fermi surface, and thus breaks the lattice point group symmetry
\cite{mackenzie2003,mazin2009,zhai2009,lee2009,raghu2009}.
As a result, orbital ordering is equivalent to the anisotropic Pomeranchuk instability of Fermi liquids.

Pomeranchuk instabilities are a large class of Fermi surface instabilities 
in the particle-hole channel with non-$s$-wave symmetry,
which can be decomposed into both density and spin-channel instabilities.
The density channel instabilities often result in uniform but 
anisotropic (nematic) electron liquid states \cite{oganesyan2001,
barci2003,lawler2006,nilsson2006, quintanilla2006, halboth2000,
dellanna2006,kee2003,yamase2005,yamase2007,kivelson2003,varma2005,
kee2005,honerkamp2005,lamas2008}.  These instabilities have been studied in the context of
doped Mott insulators~\cite{kivelson1998}, high $T_{\rm c}$ 
materials \cite{kivelson1998,kivelson2003}, and quantum Hall systems with
nearly half-filled Landau levels \cite{fradkin99,fradkin2000}.
The spin channel Pomeranchuk instabilities are a form of ``{\it unconventional 
magnetism}'' analogous to unconventional superconductivity \cite {hirsch1990, hirsch1990a, oganesyan2001,
wu2004, wu2007, varma2005, varma2006, kee2005,maslov2009}.
The instabilities result in new phases of matter, dubbed $\beta$ and $\alpha$, which respectively are counterparts to
the $B$ (isotropic) and $A$ (anisotropic) phases of $^3$He \cite{wu2004,wu2007}.
Systematic studies of the ground state properties  and collective 
excitations in both the $\alpha$ and $\beta$-phases have been performed in Refs. \ocite{wu2004} and \ocite{wu2007}.

The $t_{2g}$-orbital system of the bilayer ruthenate Sr$_3$Ru$_2$O$_7$
exhibits an unconventional anisotropic (nematic) metamagnetic state
\cite{grigera2001,grigera2004,borzi2007}, which has aroused 
much attention \cite{berridge2009,berridge2010,millis2002, green2004,tamai2008,
iwaya2007,kee2005,yamase2007,puetter2007,puetter2010}.
Sr$_3$Ru$_2$O$_7$ is a metallic itinerant system with RuO$_2$ $(ab)$ planes.
It is paramagnetic at zero magnetic field, and  below 1K develops two consecutive 
metamagnetic transitions in an external magnetic field $B$ perpendicular 
to the $ab$-plane at 7.8 and 8.1 Tesla.  Between two metamagnetic transitions, the resistivity 
measurements show a strong spontaneous in-plane anisotropy
along the $a$ and $b$-axis, with no noticeable lattice distortions.
This feature, which is presumed to be of electronic order, may be interpreted as due to nematicity resulting
from an anisotropic  distortion of the Fermi surface of the majority spin polarized by
the external magnetic field \cite{grigera2004}.
Essentially this reflects a mixture of the $d$-wave Pomeranchuk instabilities in both density and spin channels.
Recently, different microscopic theories have been constructed based 
on the quasi-1d bands of $d_{xz}$ and $d_{yz}$ by two of us 
\cite{lee2009} and also by Raghu {\it et al.} \cite{raghu2009}, and based on the 
2d-band of $d_{xy}$  by Puetter {\it et al.}\cite{puetter2010}.
In our theory, the unconventional (nematic) magnetic ordering was
interpreted as orbital ordering among the $d_{xz}$ and $d_{yz}$-orbitals.

Unlike charge and spin, orbital ordering is often difficult to measure particularly in metallic systems.
Recently, the technique of spectroscopic imaging scanning tunneling 
microscopy (SI-STM) has been applied to the active $d$-orbital 
systems of Sr$_3$Ru$_2$O$_7$\cite{lee2009qpi327} and 
Ca(Fe$_{1-x}$Co$_x$)$_2$As$_2$\cite{chuang2010}.
The SI-STM quasi-particle interference (QPI) analysis is an important
tool to study competing orders in strongly correlated systems 
\cite{kohsaka2008,wang2003qpi,balatsky2006}, and has recently been applied
to analyze the orbital band structure and orbital ordering in such systems.
The QPI pattern in Sr$_3$Ru$_2$O$_7$ exhibits characteristic square box-like
features \cite{lee2009qpi327}, and that of Ca(Fe$_{1-x}$Co$_x$)$_2$As$_2$
exhibits strong two-fold anisotropy \cite{chuang2010}.
In both cases, the QPI spectra are associated with the quasi-one dimensional $d_{xz}$ and $d_{yz}$-bands.

In a previous paper \cite{lee2009qpi}, two of us performed a theoretical analysis showing that QPI provides a sensitive method 
to detect orbital degree of freedom and orbital ordering in the quasi-1d $d_{xz}$  and $d_{yz}$ bands.
The $T$-matrix acquires momentum-dependent form factors
which extinguish certain QPI wavevectors and result in crossed
stripe features in the Fourier-transformed STM images.
The orbital ordering is reflected in the nematic distortion of the stripe QPI patterns.
These results are in qualitative agreement with recent experiments \cite{lee2009qpi327,chuang2010}.

In this article, we perform a detailed theoretical study of the QPI 
spectra in Sr$_3$Ru$_2$O$_7$ ,based on its $t_{2g}$-band structure.
Various realistic features are taken into account to construct
the tight-binding model, including the bilayer structure, the
staggered rotation of the RuO octahedra, and the on-site  spin-orbit coupling.
In addition, in order to account for the fact that STM is a surface sensitive
probe, a potential bias is added between the top and bottom layers.
Our calculation clearly shows the square box-like feature
arising from the QPI in the $d_{xz}$ and $d_{yz}$-bands,
which agrees well with the experimental data in Ref. \ocite{lee2009qpi327}.
Furthermore, we predict a reduction of the four-fold rotational ($C_4$) symmetry to two-fold ($C_2$) in the unconventional
(nematic) metamagnetic states.

This article is organized as follows.
In section \ref{sect:tightbinding}, we construct a detailed tight-binding
model to describe the bilayer $t_{2g}$-band structures.
We choose the model parameters so as to fit the experimentally measured
Fermi surface from angular resolved photon emission spectroscopy (ARPES).
In section \ref{sect:qpi}, we present the $T$-matrix method for the QPI spectra for the multi-orbital band systems.
The fact that the experimentally measured QPI is predominantly due to the top layer
is carefully taken into account.  In section \ref{sect:results}, we show the calculated QPI patterns 
and a comparison with experiments.  Predictions are then made for the QPI pattern in the presence of the
nematic orbital ordering.  Conclusions are given in section \ref{sect:conclusions}.

\section {Tight-binding model for the bilayer $t_{2g}$-orbital band}
\label{sect:tightbinding}

\begin{figure}
\vspace{10mm}
\centering\epsfig{file=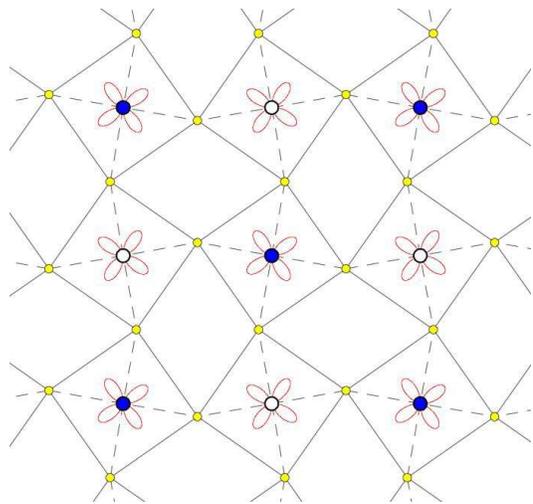,clip=1,width=0.8\linewidth,angle=0}
\caption{\label{fig:rotation} (Color online) The lattice structure in 
a single layer of Sr$_3$Ru$_2$O$_7$. 
The small yellow circle represents the octahedra oxygens which rotate
about $6.8^\circ$ (the angle in the plot is a little exaggerated) with respect to the $z$ axis on the Ru sites.
The red curves show the orientations of the Ru $d_{xy}$ orbitals.
Because the direction of the rotation is opposite for nearest neighbor
Ru sites, two types of the sublattice are identified as A (blue dot) and 
B (white dot). The direction of rotation is also opposite from bottom to 
top layers, leading to the switch of the sublattices A and B in different 
layers.}
\end{figure}

The bilayer ruthenate compound Sr$_3$Ru$_2$O$_7$ has a quasi-two dimensional
layered structure.  Its band structure in the vicinity of the Fermi level is dominated
by the $t_{2g}$-orbitals on the Ru sites, and is complicated by
the on-site spin-orbit coupling and the staggered rotation pattern of the RuO octahedra.
In this section, we derive the form of the tight-binding model based on 
symmetry considerations.
 
The lattice structure of one layer of Sr$_3$Ru$_2$O$_7$ is plotted 
in Fig. \ref{fig:rotation}, showing the rotation of the octahedra oxygens 
with opposite directions between neighboring Ru sites.
Neutron diffraction measurement\cite{shaked2000} indicated
that the rotation directions are reversed on the top and bottom layers.
This staggered rotation pattern leads to not only a unit cell doubling 
but also additional hoppings which are absent in a perfect square lattice, and
it is crucial to take this detail into account in constructing a realistic tight-binding model. 
To make the discussion simple, we divide the hopping terms into four parts: 
the in-plane hoppings existing without rotations $H_1^{\ssr INTRA}$, the
in-plane hoppings induced by the rotations $H_2^{\ssr INTRA}$,
the inter-layer hoppings existing without the rotations $H_1^{\ssr INTER}$, 
and finally the inter-layer hoppings induced by the rotations $H_2^{\ssr INTER}$.

\subsection{Uniform hopping terms without RuO octehedron rotation}
\label{subsect:unfmhop}

The Hamiltonian for $H_1^{\ssr INTRA}$ has been presented in Refs.
\ocite{raghu2009} and \ocite{puetter2010}.
Following Ref. \ocite{puetter2010}, $H_1^{\ssr INTRA}$ reads
\begin{widetext}
\bea
H_1^{\ssr INTRA}=\sum_{\vec r,s,a}
\bigg\{\hskip-0.4cm && -t_1 \Big[ \ddxz(\vec r+\xhat) \, \dnxz(\vec r) + \ddyz(\vec r+\yhat) \, \dnyz(\vec r) \Big]
-t_2 \Big[ \ddyz(\vec r+\xhat) \, \dnyz(\vec r) + \ddxz(\vec r+\yhat) \, \dnxz(\vec r) \Big] \nn \\
&& -t_3 \Big[ \ddxy(\vec r+\xhat) \, \dnxy(\vec r) + \ddxy(\vec r+\yhat) \, \dnxy(\vec r) \Big]
-t_4 \Big[ \ddxy(\vec r+\xhat+\yhat) \, \dnxy(\vec r) + \ddxy(\vec r+\xhat-\yhat) \, \dnxy(\vec r) \Big] \nn\\
&& -t_5 \Big[ \ddxy(\vec r+2\xhat) \, \dnxy(\vec r) + \ddxy(\vec r+2\yhat) \, \dnxy(\vec r) \Big] \nn \\
&&-t_6 \Big[ \ddyz(\vec r+\xhat-\yhat) \, \dnxz(\vec r) - \ddyz(\vec r+\xhat+\yhat) \, \dnxz(\vec r) \Big] \bigg\}+ {\rm H.c.} \nn \\
&&-V_{xy} \ddxy(\vec r) \dnxy(\vec r)
+  2\lambda \sum_{\vec r} \vec L(\vec r) \cdot \vec S (\vec r) \ ,
\label{eq:ham-intra1}
\eea
\end{widetext}
which includes longitudinal ($t_1$) and transverse ($t_2$) hopping for the
the $d_{xz}$ and $d_{yz}$ orbitals, respectively, as well as
are nearest neighbor ($t_3$), next-nearest neighbor ($t_4$), and next-next-nearest
neighbor ($t_5$) hopping for the $d_{xy}$ orbital. 
The summation indices $\vec{r}$, $s$, and $a$ refer to the position of Ru sites, the spin, and the layer indices.
While symmetry forbids nearest-neighbor hopping between different $t_{2g}$ orbitals 
in a perfect square lattice, due to the rotation of the oxygen octahedra, we include a term describing hopping
between $d_{xz}$ and $d_{yz}$ orbitals on next-nearest neighbor sites ($t_6$).
In each layer, the Ru sites $\rvec$ lie on a square lattice; we set the lattice constant to unity throughout.

\begin{figure}
\centering\epsfig{file=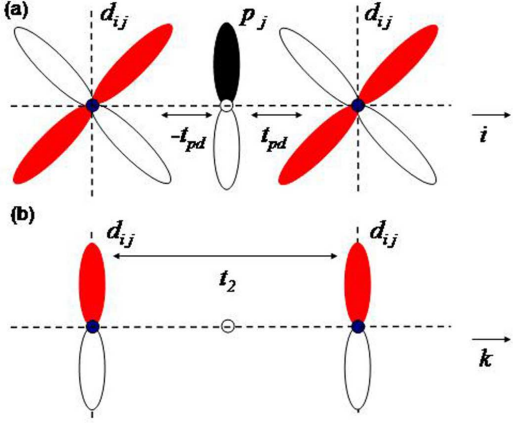,width=0.8\linewidth,clip=1,angle=0}
\caption{\label{fig:t1t2t3} (Color online) Hopping processes for 
(a) $t_1$ and $t_3$ (b) $t_2$.
For each $i,j,k$, it can be $\xhat,\yhat,\zhat$, but $i\neq j\neq k$. 
(a) The hopping processes described by $t_1$ and $t_3$ 
are assisted by the $p$ orbital of oxygens. 
(b) The hopping process described by $t_2$ is through the 
direct overlap between two identical orbitals on the 
nearest-neighbor Ru sites without going through the oxygen, thus 
it is much weaker than $t_1$ and $t_3$.
}
\end{figure}

We assume $\vert t_3\vert \approx \vert t_1\vert \gg \vert t_2\vert$, 
in accordance with the 2d nature of $d_{xy}$ and quasi-1d nature 
of $d_{yz}$ and $d_{xz}$ orbitals.  While the hopping integral $t_2$ arises from the
direct overlap of the Wannier wavefunctions for the $t_{2g}$ bands, 
the major contributions to $t_1$ and $t_3$ are from the hopping through the oxygen $2p$-orbitals. 
The corresponding hopping processes are sletched in Fig. \ref{fig:t1t2t3}.
The signs of nearest neighbor hopping integrals $t_1$ and $t_3$ can be obtained
from the second order perturbation theory:
\bea
-t_1 = {t_{pd}(-t_{pd})\over E_d-E_p} < 0.
\eea
where $t_{pd}$ is defined as the hopping integral between the ruthenium $d_{xz}$ orbital at
position $\rvec$ and the oxygen $p_z$ orbital at position $\rvec+\half\xhat$, which is identical
to that between the Ru $d_{yz}$ orbital at $\rvec$ and the O $p_z$ orbital at $\rvec+\half\xhat$.
To get $t_3$, replace the $d_{xz}$ or $d_{yz}$ orbital with the $d_{xy}$ orbital.  The sign follows from
the fact that $E_d-E_p > 0$. 
As for $t_2$, since is results from a direct overlap, as shown in Fig. \ref{fig:t1t2t3}(b), we have $t_2 > 0$.
Their magnitudes are estimated as $t_1\approx t_3 \approx 300$ meV 
and $t_2/t_1\approx 0.1$ from a fitting of LDA calculations on Sr$_2$RuO$_4$\cite{liebsch2000,eremin2002}.
For the long distance hoppings $t_{4,5}$ whose magnitudes are smaller,
their values are put by hand for later convenience.
The on-site potential for the $d_{xy}$ orbital $V_{xy}$ is introduced to 
take into account the splitting of the $d_{yz}$ and $d_{xz}$ states relative to the $d_{xy}$ states
which was found in LDA calculations \cite{singh2001}.  We take $V_{xy}/t_{1}=0.3$.

The last term in $H_1^{\ssr INTRA}$ describes the on-site spin-orbit coupling,
the energy scale of which is estimated in Ref. \ocite{haverkort2008} to be $\lambda=90$ meV,
based on a first principles study of Sr$_2$RuO$_4$.
This term couples the $d_{xy}$ and $d_{xz,yz}$ orbitals.
Truncated in the three-dimensional subspace of $t_{2g}$ orbitals
spanned by $(d_{yz}, d_{xz}, d_{xy})$, the matrix form of the $\vec L$
operators reads
\bea
L_x&=&\left(
\begin{array}{ccc}
0&0&0 \\ 0&0&i\\ 0&-i&0
\end{array}\right)
\quad,\quad
L_y=\left(
\begin{array}{ccc}
0&0&-i \\ 0&0&0\\ i&0&0
\end{array}\right)\nn\\
&&\hskip 1.0cm L_z=\left(
\begin{array}{ccc}
0&i&0 \\ -i&0&0\\ 0&0&0
\end{array}\right)\ .
\eea
It is important to notte that, unlike the usual angular momentum operators, 
the truncated matrices satisfy a different commutation relation, {\it i.e.\/}
\bea
[L_i, L_j]=-i\epsilon_{ijk} L_k.
\eea

The Hamiltonian Eq. \ref{eq:ham-intra1} is expressed in momentum
space as
\bea
H_1^{\ssr INTRA} = \sum_{\kvec,a} \Big [ \psi\yd_{s,a}(\vec k)  
\,\hat{A}_s(\kvec)\, \psi\nd_{s,a}(\kvec) + {\rm H.c.} \Big ]
\label{eq:ham-momn1}
\eea
where 
$\psi_{s,a}(\vec k)$ is defined as a 3-component spinor as
$\psi(\vec k) = \big[ d^{yz}_{s,a}(\vec k)\,,\,
d^{xz}_{s,a}(\vec k) \,,\, d^{xy}_{-s,a}(\vec k)\big]^T$ and
$d^{\alpha}_{s,a} (\vec k)$ annihilates an electron 
with orbital $\alpha$ and spin polarization $s$ at momentum $\kvec$ in the top 
($a={\rm t}$) or bottom ($a={\rm b}$) layer.   The matrix kernel $\hat{A}_s(\vec k)$ in Eq. \ref{eq:ham-momn1} is
\be
\hat{A}_s(\vec k)=
\left(\begin{array}{ccc}
\epsilon_{\kvec}^{yz}& \epsilon_{\kvec}^{\rm off}+is\lambda & ~~-s\lambda\\
\epsilon_{\kvec}^{\rm off}-is\lambda & \epsilon_{\kvec}^{xz} & ~~i\lambda\\
-s\lambda & -i \lambda &~~ \epsilon_{\kvec}^{xy}
\end{array}\right)\ ,
\ee
where the dispersions for the $d_{yz}$, $d_{xz}$, and $d_{xy}$ bands are 
\bea
\epsilon^{yz}_{\kvec} &=& - 2 t_2 \cos k_x - 2 t_1 \cos k_y, \nn \\
\epsilon^{xz}_{\kvec} &=&
  - 2 t_1 \cos k_x - 2 t_2 \cos k_y, \nn \\
  \epsilon^{xy}_{\kvec} &=& - 2 t_3 \big( \cos k_x 
  + \cos k_y \big) - 4 t_4 \cos k_x \cos k_y \nn \\
 && -2 t_5 \big( \cos 2 k_x + \cos 2 k_y \big) - V_{xy}\ ,
\eea
and
\be
\epsilon^{\rm off}_{\kvec} = - 4 t_6 \sin k_x \sin k_y.
\ee

As for $H_1^{\ssr INTER}$, since the wavefunction of the $d_{xy}$ orbital
lies largely within the $ab$-plane, its inter-layer hopping is assumed negligible
in comparison to that for the $d_{xz}$ and $d_{yz}$-orbitals.  This leads to
\be
H_1^{\ssr INTER}= -t_\perp\!\! \sum_{\alpha=xz,yz} \sum_{\kvec,s}
\Big\{ d^{\alpha\,\dagger}_{s,{\rm t}}(\kvec) \, d^{\alpha}_{s,{\rm b}}(\kvec)   + {\rm H.c.}  \Big\}
\label{hinter1}
\ee

\subsection{Staggered intra-plane hopping induced by staggered rotation of RuO octehedron}
\label{subsect:stghop}
In this subsection, we study the additional intra-plane hoppings
induced by the staggered rotation of the octahedron oxygens.
The leading effect of this rotation is to enable hopping between 
different orbitals on nearest neighbor sites.
A spin-dependent hopping between $d_{xy}$ band due to the spin-orbit coupling 
has been discussed in Ref. \ocite{fischer2010}.
In the following, we neglect the weak breaking of reflection
symmetry of each $ab$ plane due to the bilayer structure.
Since $d_{yz}$ and $d_{xz}$ are odd and $d_{xy}$ is even under this 
reflection $z\to -z$, the inter-orbital hoppings between $d_{yz}$ 
(or $d_{xz}$) and $d_{xy}$ are still zero under this assumption.
Therefore we only need to consider the hopping between $d_{yz}$ 
and $d_{xz}$ orbitals.  In the following, we will show that this 
inter-orbital hopping has
staggered signs in the real space, which causes a unit cell doubling 
as seen in LDA calculations\cite{singh2001} and ARPES 
experiment\cite{tamai2008}.

\begin{figure}
\centering\epsfig{file=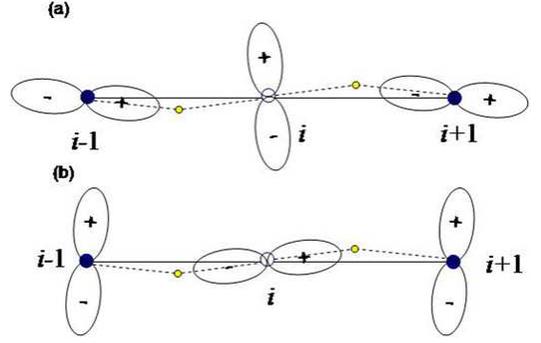,clip=1,width=0.8\linewidth,angle=0}
\caption{\label{fig:lattice} (Color online) The Wannier wavefunctions
of the $d_{yz}$ and $d_{xz}$ with the lattice distortion. 
The blue and white dots denote sublattices A (with $x+y$ odd) 
and B (with $x+y$ even), and the gray dots denote the oxygens. 
The sign indicates the sign of the wave function in the positive $z$ 
plane, and the wave functions in the negative $z$ plane have opposite signs.}
\end{figure}

We start with the hopping along the $\xhat$-direction with spin $s$
and in the layer $a$, and consider the hopping  
between $d^{yz}$ and $d^{xz}$ orbitals illustrated in Fig. \ref{fig:lattice}(a) as
\bea
-t_{\ssr{INT}}\, \big( \ddyz(\rvec) \, \dnxz(\rvec-\xhat) + {\rm H.c.}\big)\ .
\eea
This lattice structure has an inversion symmetry ${\cal I}$ with respect to site $\rvec$, 
and under such an inversion the orbitals transform as:
\bea
{\cal I} \, \dnxz(\rvec\pm\xhat) \, {\cal I}&=& \dnxz(\rvec\mp\xhat) \nn \\
{\cal I} \, \dnxz(\rvec) \, {\cal I}&=& \dnxz(\rvec)\ ,
\eea
with corresponding relations holding for the $\dnyz$ orbital.
Therefore we have
\be
{\cal I} \, \ddyz(\rvec) \, \dnxz(\rvec-\xhat) \, {\cal I} = \ddyz(\rvec) \, \dnxz(\rvec+\xhat) \ .
\label{interhopping1}
\ee
The crystal also exhibits a reflection symmetry with respect to the $yz$ planes containing the oxygen sites. 
Let us define ${\cal J}$ as the reflection operation with respect to the $yz$ 
plane containing the oxygen site between $\rvec$ and $\rvec+\xhat$.
Under the operation of ${\cal J}$,
\bea
{\cal J} \, \dnxz(\rvec) \, {\cal J} &=& -  \dnxz(\rvec+\xhat) \nn \\
{\cal J} \, \dnyz(\rvec) \, {\cal J} &=& + \dnyz(\rvec+\xhat) \ .
\eea
Thus,
\be
{\cal J} \, \ddyz(\rvec) \, \dnxz(\rvec+\xhat) \, {\cal J} = -\ddyz(\rvec+\xhat) \, \dnxz(\rvec) \ ,
\label{interhopping2}
\ee
Combining Eq.s \ref{interhopping1} and \ref{interhopping2} leads to:
\be
{\cal JI} \, \ddyz(\rvec) \, \dnxz(\rvec-\xhat) \, {\cal IJ} = -\ddyz(\rvec+\xhat) \, \dnxz(\rvec) \ ,
\ee
which means that this hopping is {\it staggered\/}.

Note that the above discussion is generally valid regardless of the 
intermediate state of the hopping process.
The intermediate state, however, is important to give the second
order perturbation expression for $t_{\ssr{INT}}$ as
\bea
t_{\rvec\rvec'}^{\alpha\beta} = -\sum_m {\langle \rvec,\alpha\vert H_{\rm RuO} \vert 
m\rangle\langle m\vert H_{\rm RuO} \vert \rvecp,\beta\rangle \over E_d-E_m},
\eea
where $\alpha,\beta=xz,yz$.  $H_{\rm RuO}$ describes the hopping between 
the $t_{2g}$ orbital on Ru sites and the $2p$ orbitals on neighboring 
O sites. $\vert m\rangle$ denotes an oxygen $2p$ orbital, which is
an intermediate state for the Ru-Ru hopping processes.
Because of the reflection symmetry with respect to the $xy$ plane and 
the fact that $d_{yz}$ and $d_{xz}$ are odd under this reflection, 
$\langle \rvec,\alpha\vert H_{\rm RuO} \vert m\rangle$ is non-zero only 
if the intermediate state is also odd under this reflection. 
As a result, $\vert m\rangle$ can be only $\vert p_z\rangle$.
However, in order to determine the sign and the magnitude of 
$t_{\ssr{INT}}$, a detailed knowledge of the pseudopotentials for the Hamiltonian $H_{\rm RuO}$ 
is required, which is beyond the scope of this paper. 
Nevertheless, since this term is expected to be small and its 
main consequence is to provide the necessary coupling between $\kvec$ and $\kvec+\Qvec$, 
where $\Qvec=(\pi,\pi)$, we can treat it as a fitting parameter.

Similar reasoning can be applied for the hybridized hopping
between $d_{xz}$ and $d_{yz}$-orbitals along the $\hat y$-direction, 
which is also staggered.
Furthermore, the $C_4$ symmetry around each Ru site
relates the staggered hoppings along the $\xhat$ and $\yhat$-directions.
Putting all the above together, we arrive at the staggered in-plane hopping
contribution to the Hamiltonian 
\begin{widetext}
\be
H_2^{\ssr INTRA} = -t_{\ssr{INT}} \sum_{\rvec,s,a,{\hat\delta}}
(-)^a \, e^{i\Qvec\cdot\rvec} \, \Big[ \ddyz(\rvec) \, \dnxz(\rvec+{\hat\delta}) - \ddxz(\rvec) \, \dnyz(\rvec+{\hat\delta}) \Big]
 + {\rm H.c.}\ ,
\label{eq:hopping-intra2}
\ee
\end{widetext}
where ${\hat\delta}$ ranges over $\xhat$ and $\yhat$, $(-1)^a=\mp 1$ for top and bottom layers, respectively,
and where in our convention $e^{i\Qvec\cdot\rvec}=\mp 1$ for $\rvec$ in the A (B) sublattice.
Note that there is only a single independent parameter $t_{\ssr{INT}}$ to characterize this in-plane staggered hopping.

It is straightforward to transform Eq. \ref{eq:hopping-intra2} into
momentum space as
\begin{widetext}
\be
H_2^{\ssr INTRA}
= -2t_{\ssr{INT}}{\sum_{\kvec,s,a}}^\prime  (-)^a\,(\cos k_x+\cos k_y)
\Big[ \ddyz(\kvec+\Qvec) \, \dnxz(\kvec) - \ddxz(\kvec+\Qvec) \, \dnyz(\kvec) \Big] + {\rm H.c.}\ ,
\label{hintra2}
\ee
\end{widetext}
where the prime on the sum indicates that $\kvec$ is restricted to only half of the Brillouin zone.

\subsection{Inter-layer staggered hopping}
In this subsection, we study the additional hybridized inter-layer hopping
between different orbitals, {\it i.e.\/}, the $H_2^{\ssr INTER}$ term.
This contribution arises because the rotation patterns of the RuO octahedra in the
two layers are opposite to each other.  Because the $d_{xy}$ and $d_{xz/yz}$ orbitals
have different azimuthal quantum number of orbital angular momentum, they do not
mix, even in the presence of the RuO octahedra rotation.
The leading order inter-layer hybridization therefore occurs between $d_{xz}$ and $d_{yz}$ orbitals,
and the hybridization Hamiltonian is
\bea
H_2^{\ssr INTER} &=& - \sum_\rvec e^{i\Qvec\cdot\rvec}
\Big[t_{\rm bt}^{(1)} \, d^{yz\,\dagger}_{s,{\rm t}}(\rvec) \, d^{xz}_{s,{\rm b}}(\rvec) \\
&&\qquad\qquad +t_{\rm bt}^{(2)} d^{xz\,\dagger}_{s,{\rm t}}(\rvec)d^{yz}_{s,{\rm b}}(\rvec) \Big]  
+ {\rm H.c.}\ . \nn
\eea

\begin{figure}
\includegraphics[width=0.7\linewidth]{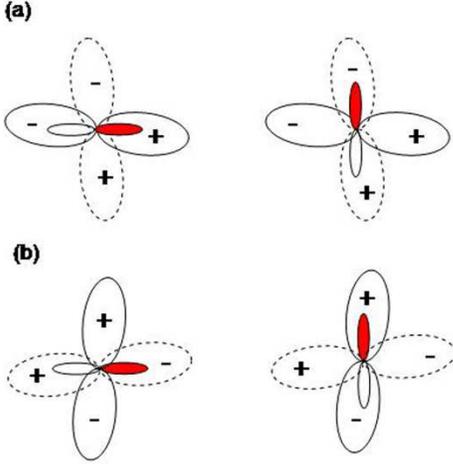}
\caption{\label{fig:inter} (Color online) The wave functions viewed from the top of the material.  The dashed line represents the wave function
of the $d$ orbitals on the top layer, and the solid line for those on the bottom layer.  The smaller figures represent the $p$ orbital of the 
oxygen between layers with the red lobe having positive sign and the white lobe having negative sign. Note that the signs of the $d$
orbitals indicates those of the wave functions closest to the the oxygen.  (a) $d_{xz}$ at bottom layer and $d_{yz}$ at top layer, and
(b) $d_{yz}$ at bottom layer and $d_{xz}$ at top layer.}
\end{figure}

Next we use the second order perturbation theory to derive
the staggered inter-layer hopping intergrals.
We consider two hopping processess: 
(1) hopping from $d_{xz}$ orbital at sublattice A on the
bottom layer to $d_{yz}$ orbital at sublattice B on the top layer,
and (2) hopping from $d_{yz}$ orbital at sublattice A on
the bottom layer to $d_{xz}$ orbital at sublattice B on the top layer.
The hopping intgrals for these two processes can be written as:
\bea
t_{\rm bt}^{(1)} = -\sum_m {\langle \rvec,yz,{\rm b}\vert H_{\rm RuO}
\vert m\rangle\langle m\vert H_{\rm RuO} \vert \rvec,xz,{\rm t}\rangle \over E_d-E_m},\nn\\
t_{\rm bt}^{(2)} = -\sum_m {\langle \rvec,xz,{\rm b}\vert H_{\rm RuO}
\vert m\rangle\langle m\vert H_{\rm RuO}\vert \rvec,yz,{\rm t}\rangle \over E_d-E_m}
\eea
where $i$ belongs to sublattice A in the bottom layer and sublattice
B in the top layer by our convention. 
Because the $d_{xz}$ and $d_{yz}$ are odd under the rotation 
of $90^\circ$ with respect to the $z$ axis despite of the O-octahedral 
rotation, their overlaps with $p_z$ are zero. 
Therefore the these two processes can only go through $p_x$ and $p_y$ 
orbitals of the oxygen between the layers. 
Fig. \ref{fig:inter} presents the views of wavefunctions from the topview.
It should be noted that for the top layer, the components of the wave 
functions having largest overlap with the oxyegn $p$ orbitals are
the one in the negative $z$ so that there is an additional minus
sign in addition to those plotted in the Fig. \ref{fig:lattice}.
Unlike the case of $t_{\ssr{INT}}$, because the Ru atoms on the top and bottom layers and the oxygen 
between them are colinear,
the signs of $t^{1,2}_{bt}$ can be determined from the geometry shown in Fig. \ref{fig:inter}.
We can then obtain:
\bea
&&\langle \rvec,xz,{\rm b}\vert H_{\rm RuO}\vert p_x,0\rangle \cdot
\langle p_x,0\vert H_{\rm RuO}\vert \rvec,yz,{\rm t}\rangle > 0\nn\\
&&\langle \rvec,xz,{\rm b}\vert H_{\rm RuO}\vert p_y,0\rangle \cdot
\langle p_y,0\vert H_{\rm RuO}\vert \rvec,yz,{\rm t}\rangle > 0\nn\\
&&\langle \rvec,yz,{\rm b}\vert H_{\rm RuO}\vert p_x,0\rangle \cdot
\langle p_x,0\vert H_{\rm RuO}\vert \rvec,xz,{\rm t}\rangle < 0\nn\\
&&\langle \rvec,yz,{\rm b}\vert H_{\rm RuO}\vert p_y,0\rangle \cdot
\langle p_y,0\vert H_{\rm RuO}\vert \rvec,xz,{\rm t}\rangle < 0\nn \ ,
\eea
where $\vert p_x,0\rangle$ is the oxygen $2p_x$ orbital with planar position $\rvec$ situated
midway between the top (t) and bottom (b) ruthenium sites.
Together with $E_d-E_p>0$, we conclude that 
$t_{\rm bt}^{(1)} = - t_{\rm bt}^{(2)} \equiv t^\perp_{\ssr{INT}} > 0$. 
It can also be easily generalized that if $\rvec$ belongs sublattice 
B (A) in the bottom (top) layer, we 
have obtain the same result except an opposite sign.

Now we transform into momentum space, after which the $H_2^{\ssr INTER}$ term reads
\begin{widetext}
\be
H_2^{\ssr INTER}
= -t^\perp_{\ssr{INT}} {\sum_{\kvec}}' \Big[d^{yz\, \dagger}_{s,{\rm t}}(\kvec+\Qvec) \, d^{xz}_{s,{\rm b}}(\kvec)
- d^{xz\, \dagger}_{s,{\rm t}}(\kvec+\Qvec) \, d^{yz}_{s,{\rm b}}(\kvec)
+ d^{xz\, \dagger}_{s,{\rm b}}(\kvec+\Qvec) \, d^{yz}_{s,{\rm t}}(\kvec)
- d^{yz\, \dagger}_{s,{\rm b}}(\kvec+\Qvec) \, d^{xz}_{s,{\rm t}}(\kvec) \Big] + {\rm H.c.}
\label{hinter2}
\ee
\end{widetext}

\subsection{Fermi surfaces}
\begin{figure*}
\includegraphics[width=0.8\linewidth]{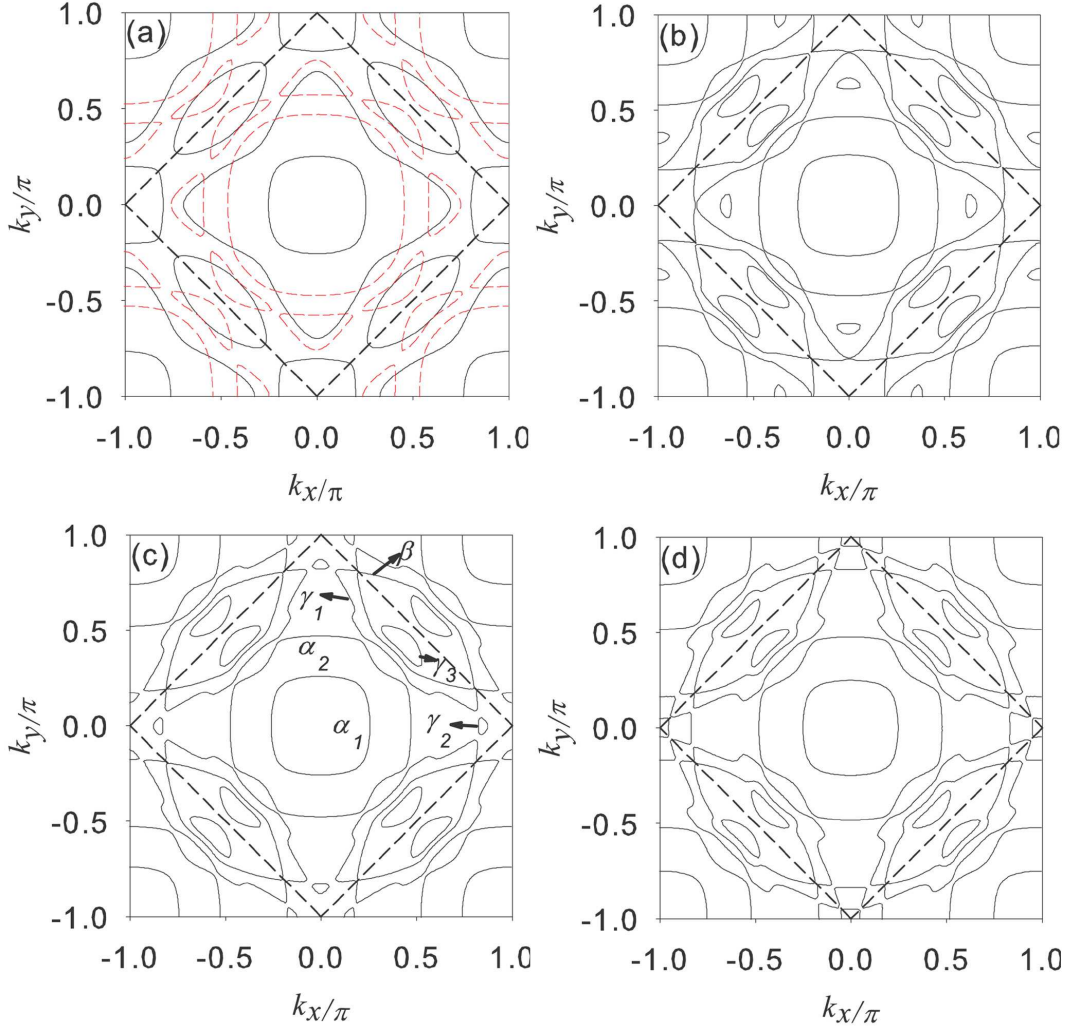}
\caption{\label{fig:fs} (Color online) The Fermi surfaces using the 
bilayer tight-binding model with the parameters: $t_{1} = 0.5, 
t_{2} = 0.05, t_{3} = 0.5, t_{4} = 0.1,t_{5} = -0.03, t_{6} = 0.05,
t_\perp=0.3, t_{\ssr{INT}}=t^\perp_{\ssr{INT}}=0.05, \lambda = 0.1, V_{xy}=0.15$, 
and $\mu=0.47$ for (a) $V_{\rm bias}=0$, (b) $V_{\rm bias}=0.1$, (c) $V_{\rm bias}=0.2$, and 
(d) $V_{\rm bias}=0.3$. 
The thick dashed lines mark the boundary of half Brillouin zone due
to the unit cell doupling induced by the rotation of octahedra oxygens. 
(a) For $V_{\rm bias}=0$, the Fermi surfaces of the bonding ($k_z=0$, black 
solid lines) and the anti-bonding bands ($k_z=\pi$, red dashed lines) 
could cross since $k_z$ is a good quantum number.
(b) As $V_{\rm bias}$ is turned on, the crossings of the Fermi surfaces 
with different $k_z$ are avoided.
(c) The optimized The Fermi surfaces are obtained with $V_{\rm bias}=0.2$. 
Fermi surface shhets of $\alpha_1$,  $\alpha_2$, $\gamma_1$, $\gamma_2$, $\gamma_3$, 
and $\beta$ are marked. 
The $\gamma_{1,2}$ sheets have dominant 2-D $d_{xy}$ orbital character 
while the $\alpha_{1,2}$ sheets are mostly formed by quasi-1d 
$d_{yz,xz}$ orbitals.
The $\gamma_3$ sheets are not seen in the ARPES measurements.
(d) For $V_{\rm bias}=0.3$, the Fermi sheets of $\gamma_2$ disappear.
}
\end{figure*}

Adding up the contributions from Eqs. \ref{eq:ham-momn1}, \ref{hinter1}, \ref{hintra2}, 
and \ref{hinter2} leads to the tight-binding model
\bea
H_0&=&H_1^{\ssr INTRA}+H_1^{\ssr INTER}+H_2^{\ssr INTRA}+H_2^{\ssr INTER} \\
&=& {\sum_{\kvec}}' \phi\yd_{\kvec,s} \, {\cal H}\nd_\kvec \> \phi\nd_{\kvec,s}\ ,\nn
\eea
where
\be
 {\cal H}_\kvec=
\left( \begin{array}{cccc}
\hat{L}^+_s(\kvec)&-\hat{G}^\dagger(\kvec)&\hat{B}_1^\dagger&\hat{B}^\dagger_2\\
-\hat{G}(\kvec)&\hat{L}^+_s(\kvec+\Qvec)&\hat{B}^\dagger_2&\hat{B}^\dagger_1\\
\hat{B}_1&\hat{B}_2&\hat{L}^-_s(\kvec)&\hat{G}^\dagger(\kvec)\\
\hat{B}_2&\hat{B}_1&\hat{G}(\kvec)&\hat{L}^-_s(\kvec+\Qvec) 
\end{array} \right)\ .
\label{eq:h0}
\ee
and
\be
\phi^{\dagger}_{\kvec,s}=\big(\psi^{\dagger}_{s,{\rm t}}(\kvec) \,,\,
\psi^{\dagger}_{s,{\rm t}}(\kvec+\Qvec) \,,\, \psi^{\dagger}_{s,{\rm b}}(\kvec) \,,\,
\psi^{\dagger}_{s,{\rm b}}(\kvec+\Qvec)\big)\ ,
\label{phidag}
\ee
with $\psi\yd_{s,a}(\vec k) = \big( \ddyz(\vec k)\,,\,\ddxz(\vec k)\,,\, d^{xy\,\dagger}_{-s,a}(\vec k)\big)$
as before (see Eq. \ref{eq:ham-momn1}).  The matrix kernels 
$\hat{L}^a_s(\kvec)$, $\hat{G}(\kvec)$, $\hat{B}_1$, and $\hat{B}_2$ 
in Eq. \ref{eq:h0} are defined as
\be
\hat{L}^a_s(\kvec)=\hat{A}_s(\kvec) - \big(\mu- \half (-1)^a\,V_{\rm bias}\big) \, {\hat I}\ ,
\ee
\be
\hat{G}(\kvec)= \left(
\begin{array}{ccc}
0&-2t_{\ssr{INT}}\,\gamma(\vec k)&0\\
2t_{\ssr{INT}}\,\gamma(\vec k) &0&0\\
0&0&0\end{array}\right)\ ,
\ee
and
\be
\hat{B}_1=\left(\begin{array}{ccc}
-t_\perp&0&0\\ 0&-t_\perp&0\\ 0&0&0
\end{array} \right) \quad,\quad
\hat{B}_2 =\left(
\begin{array}{ccc}
0&t^\perp_{\ssr{INT}}&0\\
-t^\perp_{\ssr{INT}}&0&0\\
0&0&0
\end{array}\right)\ ,
\ee
where $\gamma(\vec k)=\cos k_x +\cos k_y$,
$\mu$ is the chemical potential, and 
$V_{\rm bias}$ is the difference of on-site potential in the top and bottom RuO layers.
The $V_{\rm bias}$ term induces more splitting of bonding and anti-bonding 
solutions between layers, as will be discussed in the following sections.

For $V_{\rm bias}=0$, $H_0$ can be reduced to two independent parts classified 
by the bonding and anti-bonding solutions with respect to the layers.
To see this, first we perform a gauge transformation in $H_0$, sending
$d^{yz,xz}_{s,{\rm b}}(\kvec+\Qvec)\to - d^{yz,xz}_{s,{\rm b}}(\kvec+\Qvec)$.
Then we introduce $k_z=0,\pi$ to perform a Fourier transform on 
the layer index. 
We have:
\bea
H_0(V_{\rm bias}=0) &=& h_0(k_z=0) + h_0(k_z=\pi), 
\eea
with $h_0(k_z)$ defined as
\be
h_0(k_z) = {\sum_\kvec}' \Phi^{\dagger}_{\kvec,s,k_z}
\left(
\begin{array}{cc}
\hat{h}_{0s}(\kvec,k_z)& \hat{g}^\dagger(\kvec,k_z)\\
\hat{g}(\kvec,k_z)&\hat{h}_{0s}(\kvec+\Qvec,k_z)\\
\end{array}\right) \Phi_{\kvec,s,k_z}\ .
\label{eq:h0kz}
\ee
In Eq. \ref{eq:h0kz},
 $\hat{h}_{0s}$, $\hat{g}(\kvec,k_z)$ and $\Phi^{\dagger}_{\kvec,s,k_z}$
are defined as
\bea
\hat{h}_{0s}(\kvec,k_z) &=& \hat{A}_s(\kvec) + \hat{B}_1\cos k_z \\
\hat{g}(\kvec,k_z) &=& \hat{G}(\kvec) - 2\hat{B}_2 \cos k_z \nn 
\eea
and
\bea
\Phi^{\dagger}_{\kvec,s,k_z}
&=&\big( d^{yz\, \dagger}_{\kvec,s,k_z} \,,\,
d^{xz\,\dagger}_{\kvec,s,k_z} \,,\,
d^{xy\,\dagger}_{\kvec,-s,k_z} \,,\, \\
&& \qquad  d^{yz\,\dagger}_{\kvec+\Qvec,s,k_z} \,,\, d^{xz\,\dagger}_{\kvec+\Qvec,s,k_z} \,,\,
d^{xy\,\dagger}_{\kvec+\Qvec,-s,k_z} \big) \ . \nn
\eea

\begin{figure}
\includegraphics[width=0.9\linewidth]{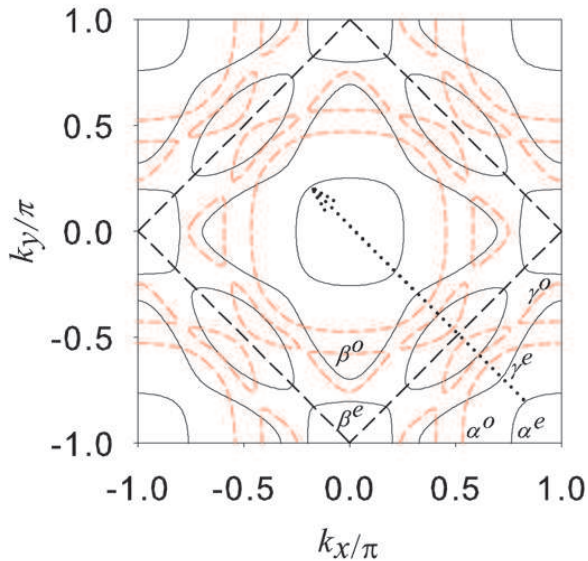}
\caption{\label{fig:fs-rep} (Color online) The analysis of the Fermi surface formation for $V_{\rm bias}=0$.
Two copies of the Fermi surfaces of Sr$_2$RuO$_4$
are labeled as $\alpha^e$, $\beta^e$, $\gamma^e$ for bonding and $\alpha^o$, $\beta^o$, $\gamma^o$ for anti-bonding bands.
The back-folding of the Brillouin zone from the corners produces identical partners for each band appearing at positions connected by the 
wave vector $\Qvec=(\pi,\pi)$ (the dotted arrow), leading to the Fermi surfaces plotted in Fig. \ref{fig:fs}(a).}
\end{figure}

The Fermi surface for $V_{\rm bias}=0$ is plotted in Fig. \ref{fig:fs}(a).  It consists of many disconnected sheets.
Since $k_z$ is a good quantum number, the individual Fermi surfaces of the 
bonding and anti-bonding bands could cross; this in fact
makes it easier to analyze how the Fermi surfaces are formed due 
to hybridization among the $t_{2g}$ bands.  It has been illustrated in Ref. \ocite{mercure2009} that the 
Fermi surface of Sr$_3$Ru$_2$O$_7$ can be schematically understood from that 
of Sr$_2$RuO$_4$.  In Sr$_2$RuO$_4$, the hybridizations of the $t_{2g}$ bands result in three eigenbands:
$\alpha$ and $\beta$ bands with mostly quasi-1d $d_{yz}$ and $d_{xz}$ 
characters, and $\gamma$ band with dominant $d_{xy}$ character.
For $V_{\rm bias}=0$, we can begin from two copies of the Fermi surfaces of Sr$_2$RuO$_4$ since the bilayer splitting doubles for each band.
From our calculations, three bonding bands ($\alpha^e$, $\beta^e$, and $\gamma^e$) and three antibonding bands
($\alpha^o$, $\beta^o$, and $\gamma^o$) are clearly identified, as shown in Fig. \ref{fig:fs-rep}.
Finally, due to the unit cell doubling induced by the rotated oxygen octahedra, 
the Brillouin zone is back-folded from the corners with respect to the dashed lines.
As a result, each of the six bands will have an identical partner appearing at positions connected by the wave vector $\Qvec=(\pi,\pi)$,
producing the Fermi surfaces plotted in Fig. \ref{fig:fs}(a).

When $V_{\rm bias}\neq 0$, the crossings of the Fermi surfaces between bond and anti-bonding bands can be avoided
because the $V_{\rm bias}$ term breaks the bilayer symmetry.  To match the observed ARPES results\cite{tamai2008},
it is crucial to avoid these crossings in order to obtain the 
correct shapes of the Fermi surface sheets.  This suggests that a finite $V_{\rm bias}$ is a necessary aspect of any realistic model.
Figs. \ref{fig:fs}(b)-(d) show the Fermi surfaces with several different value of $V_{\rm bias}$, and it can be seen that 
the crossings of the Fermi surfaces are all avoided when $V_{\rm bias}\ne 0$.
Fig. \ref{fig:fs}(c) shows the Fermi surface with optimized parameters fit to the ARPES experiment\cite{tamai2008}.
The agreement with experiment appears satisfactory.  The Fermi surfaces of $\alpha_1$,  $\alpha_2$, $\gamma_1$, $\gamma_2$,
and $\beta$ identified from the ARPES are clearly reproduced with the correct shapes.  Moreover, the average filling per Ru atom
with these optimized parameters is 4.05, which is also consistent with the valence charge of Ru atoms in Sr$_3$Ru$_2$O$_7$.

One major discrepancy is the appearance of additional electron Fermi pockets, $\gamma_3$,
enclosed by the $\beta$ bands as shown in Fig. \ref{fig:fs}(c).  
While the LDA calculation also showed the existence of $\gamma_3$ pockets, ARPES did not observe them. 
We suspect that this band might be too small to be resolved in the spectral weight measured by ARPES, and 
other measurements like quantum oscillations might be more sensitive to this band.

\section{$T$-matrix formalism for the multiband systems}
\label{sect:qpi}

QPI imaging has been studied using a $T$-matrix formalism for various systems including the high-$T_{\rm c}$ cuprates\cite{wang2003qpi,balatsky2006}, 
multiband systems with quasi-1d $d$-bands\cite{lee2009qpi}, iron-pnictide superconductors\cite{zhang2009qpi}, 
and topological insulators Bi$_2$Te$_3$\cite{lee2009tiqpi,zhou2009}, {\it etc.\/}.   The scattering mechanism for the quasiparticles is
usually taken to be elastic impurities, and is modeled by a local variation of the orbital energies.  
Because the impurites are introduced mainly on the surface of the material\cite{lee2009qpi327},
we consider a single impurity at $\rvec=0$ on the top layer only. 
Assuming that the impurity has the same effect for all orbitals, the impurity potential is modeled by
\be
H^{\ssr{IMP}}=V_0\sum_\alpha d^{\alpha\,\dagger}_{s,{\rm t}}(\rvec=0)\,d^\alpha_{s,{\rm t}}(\rvec=0)\ ,
\ee
where the orbital label $\alpha$ runs over all three possibilities $xy$, $yz$, and $xz$.
In Fourier space, then,
\bea
H^{\ssr{IMP}}&=&{V_0\over N}\sum_{\kvec,\kvec',\alpha} d^{\alpha\,\dagger}_{s,{\rm t}}(\kvec)\,d^\alpha_{s,{\rm t}}(\kvec') \nn \\
&=&{1\over N}{\sum_{\kvec,\kvec',s}}' \phi\yd_{\kvec,s}\,{\hat V}\,\phi\nd_{\kvec's}\ ,
\eea 
where
\be
{\hat V}=\left(\begin{array}{cc}
V_0\hat{M}&{\mathbb O}\\{\mathbb O}&{\mathbb O}\end{array}\right) \quad,\quad
{\hat M}=\left(\begin{array}{cc}
{\hat I}&{\hat I}\\{\hat I}&{\hat I}\end{array}\right)\ ,
\ee
with ${\hat I}$ the $3\times 3$ identity matrix, ${\mathbb O}$ is a $6\times 6$ matrix of zeroes,
and where $\phi\yd_{\kvec,s}$ is defined in Eq. \ref{phidag}.

Extending the standard $T$-matrix formalism to multiband systems\cite{lee2009qpi}, we have that
the Green's function satisfies the following matrix equation with dimension $12\times 12$:
\be
\hat{G}(\kvec,\vec{p},\omega) = \hat{G}^0(\kvec)\,\delta_{\kvec,\vec{p}} + \hat{G}^0(\kvec)\,\hat{T}(\kvec,\vec{p},\omega)\,\hat{G}^0(\vec{p})\ ,
\label{green}
\ee
where $\hat{G}^0(\kvec)$ is the unperturbed Green's function defined as:
\be
\hat{G}^0(\kvec) = \left[\omega + i\eta -\hat{H}_0(\kvec)\right]^{-1}
\ee
and $\hat{T}(\kvec,\vec{p},\omega)$ is the $T$-matrix, which satisfies
\be
\hat{T}(\kvec,\vec{p},\omega) = \hat{V}(\kvec,\vec{p}) +{1\over N}{\sum_{\kvec'}}'
\hat{V}(\kvec,\kvec') \, \hat{G}^0(\kvec') \, \hat{T}(\kvec',\vec{p},\omega)
\ee
Note that the momenta $\kvec$ and $\vec{p}$ are both restricted the half Brillouin zone.
Since $\hat{V}$ is momentum-independent, the T-matrix is also momentum-independent which can be easily evaluated as:
\be
\hat{T}(\omega) = \Bigg[\hat{I} - \hat{V}\bigg({1\over N}{\sum_{\kvec'}}' \hat{G}^0(\kvec')\bigg)\Bigg]^{-1}\!\!\hat{V}\ .
\ee
The local density of states (LDOS) on the layer $a$ for orbital $\alpha$, spin $s$ at position $\vec{r}$ and sample bias voltage $V$, 
$\rho^\alpha_{s,a}(\vec{r},E=eV)$ can be evaluated by:
\bea
\rho^\alpha_{s,a}(\vec{r},E)&=&{1\over N}{\sum_{\kvec,\vec{p}}}' e^{i(\vec{p}-\kvec)\cdot\vec{r}}
\Big[ {\cal G}^\alpha_{s,a}(\kvec,\vec{p},E) \nn \\
&&\hskip1.0cm + {\cal G}^\alpha_{s,a}(\kvec+\Qvec,\vec{p}+\Qvec,E)\Big]\nn\\
&& +e^{i(\vec{p}-\kvec-\Qvec)\cdot\vec{r}}
\Big[{\cal G}^\alpha_{s,a}(\kvec+\Qvec,\vec{p},E) \nn \\
&&\hskip1.0cm + {\cal G}^\alpha_{s,a}(\kvec,\vec{p}+\Qvec,E)\Big]\ ,
\label{ldos}
\eea
where ${\cal G}^\alpha_{s,a}(\kvec,\kvec',\omega)=\int \! dt \> e^{i\omega t}\langle T_t\, d^\alpha_{\kvec,s,a}(t) d^{\alpha\,\dagger}_{\kvec',s,a}(0)\rangle$ 
can be read off from Eq. \ref{green}. 
Generally speaking, the differential conductance $dI/dV$ measured by the STM is proportional to the LDOS.
However, special care must be taken in order to account for certain experimental details, as we will discuss in the following section.

\section{Results}
\label{sect:results}

\subsection{General discussions}
First, it is important to mention that because experimentally the tip of the STM is much closer
to the top layer, it predominantly measures the LDOS on the top layer.  Second, because the wave functions for different orbitals
could have different overlaps with the STM tip, the tunneling matrix elements may be orbital-dependent.
Therefore, the simplest model to relate the conductance $dI/dV$ and the corresponding LDOS can be written as: 
\be
\frac{dI}{dV}(\vec{r},E)\propto \rho(\vec{r},E)\equiv \sum_{\alpha,s}C^{\alpha}\rho^\alpha_{s,{\rm t}}(\vec{r},E).
\ee
Finally, the QPI imaging can be obtained by performing the Fourier transformation of $\rho(\vec{r},E)$, {\it viz.\/}
\be
\rho(\vec{q},E)={1\over N}\sum_{\vec{r}} e^{-i\vec{q}\cdot\vec{r}}\rho(\vec{r},E) \ .
\ee
In this paper, we plot $\vert \rho(\vec{q},E)\vert$ only for $\vec{q}\neq 0$ since we are interested only in the change of the local density of states
due to the impurity.  A $101\times 101$ square lattice is used in the wavevector summations, and a broadening factor $\eta=0.02$ ({\it i.e.\/}
an imaginary part to the energy) is introduced by hand.

\begin{figure}
\includegraphics[width=0.9\linewidth]{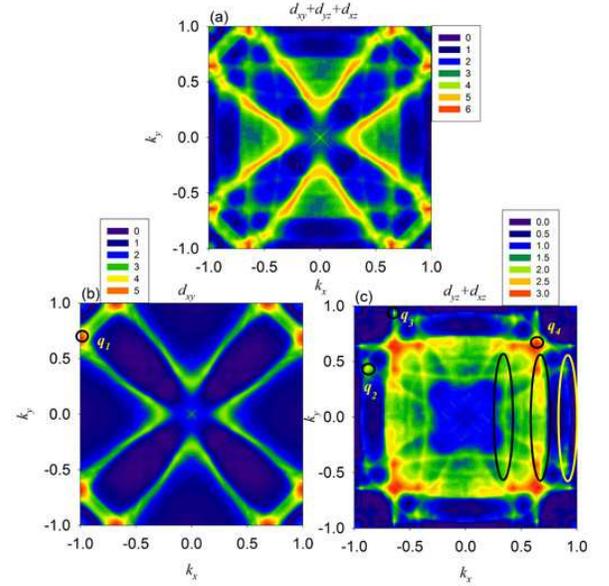}
\caption{\label{fig:qpi} (Color online) QPI imaging at zero sample bias voltage ($E=0$) contributed from scatterings 
(a) within all $t_{2g}$ bands, (b) within 2-D band $d_{xy}$, and (c) within two quasi-1d bands $d_{yz}$ and $d_{xz}$.
The scattering potential is introduced only for top and bottom layers with $V_0=1.0$, reflecting the fact that the impurities are usually in the 
top layer. Only the LDOS on the top layer are calculated.
(b) The strong features are due to the scatterings between the small hole pockets $\gamma_2$ and the parts of $\gamma_1$ marked by the solid lines 
in Fig. \ref{fig:qpi-sc}. A representative strongest wavevector $\vec{q}_1$ is also indicated.
(c) The strongest wavevectors $\vec{q}_{2-4}$ can be understood by scatterings indicated in Fig. \ref{fig:qpi-sc}.
The stripe-like features enclosed by the ovals (both black and yellow) result from the flat parts of the $\alpha_{1,2}$ bands, which are the signatures of the quasi-1d bands. 
}
\end{figure}

\begin{figure}
\centering\epsfig{file=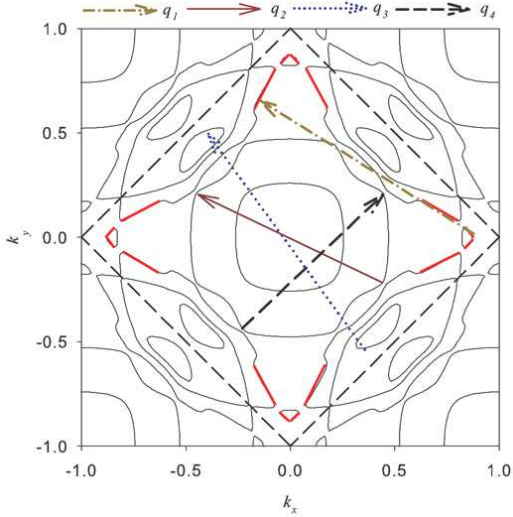,clip=1,width=0.8\linewidth,angle=0}
\caption{\label{fig:qpi-sc} (Color online) The scattering processes related to strongest features in Fig. \ref{fig:qpi}. 
The scatterings within and between the parts of Fermi surfaces marked by the red solid lines, which are mostly from $\gamma_1$ and $\gamma_2$ pcoktes, 
contribute the dorminant features in the QPI image of 2d band $d_{xy}$ shown in Fig. \ref{fig:qpi}(b).
A representative strongest wavevector $\vec{q}_1$ shown in Fig. \ref{fig:qpi}(b) is plotted.
As for the QPI image of quasi-1d bands $d_{yz,xz}$ shown in Fig. \ref{fig:qpi}(c), 
the dorminant scatterings related to strongest wavevectors $\vec{q}_{2-4}$ occur mostly within $\alpha_1$ band, as indicated by the arrows.
}
\end{figure}

We first compute the QPI imaging at zero sample bias voltage ($E=0$).
Fig. \ref{fig:qpi}(a) shows the QPI imaging due to impurity scattering from all three $t_{2g}$ bands.  The plot exhibits
several features which can be understood as follows.
Since the contributions to the LDOS from different $t_{2g}$ bands can be computed independently as seen in Eq. \ref{ldos}, 
we also compute separately the QPI imaging for the 2-D $d_{xy}$ band (Fig. \ref{fig:qpi}(b)) and for the quasi-1d $d_{yz}$ and $d_{xz}$
bands (Fig. \ref{fig:qpi}(c)) for comparison.
The strong features seen in Fig. \ref{fig:qpi}(b) come from the scatterings within and between $\gamma_{1,2}$ pockets (the red solid lines in 
Fig. \ref{fig:qpi-sc}).  This is to be expected, since both pockets have dominant $d_{xy}$ orbital character.
As for Fig. \ref{fig:qpi}(c), the signature stripe-like patterns of the quasi-1d bands\cite{lee2009qpi} can clearly be seen, 
and we find that the dominant features largely come from the $\alpha_2$ band scatterings, as indicated in Fig. \ref{fig:qpi-sc}.
The reason why the $\alpha_2$ band scatterings are much more prominent than the $\alpha_1$ band scatterings is that  the
$\alpha_2$ ($\alpha_1$) band is mostly composed of the anti-bonding (bonding) solution with respect to the layers,
with more (less) weights on the top layer.   Since we only compute the LDOS on the top layer, the $\alpha_2$ band
scatterings are much more important than the $\alpha_1$ band scatterings.

Another general feature present in Figs. \ref{fig:qpi}, also \ref{fig:qpi-dispersion} and \ref{fig:qpi-real} is that while 
the Fermi surfaces without a nematic order have not only the $C_4$ symmetry but also inversion symmetries with respect to $k_x$ and $k_y$ axes, 
the QPI patterns do not have the inversion symmetries with respect to $q_x$ and $q_y$ axes.
The reason for this discrepancy is delicate, and we will explain in the following.
As can be seen in Fig. \ref{fig:rotation}, the inversion symmetry is defined only as the inversion axis chosen to pass through the oxygen sites.
Since we have the degree of freedom to choose the inversion axis as computing the Fermi surfaces, Bloch theorem ensures that the system has the inversion symmetry. 
However, when computing QPI patterns, we have to put an impurity on one Ru site. 
As a result, we can only choose the inversion axis passing through this impurity at Ru site, which explicitly breaks the inversion symmetries. 
This explains why the QPI patterns do not have the inversion symmetries as the Fermi surfaces do.

\begin{figure}
\centering\epsfig{file=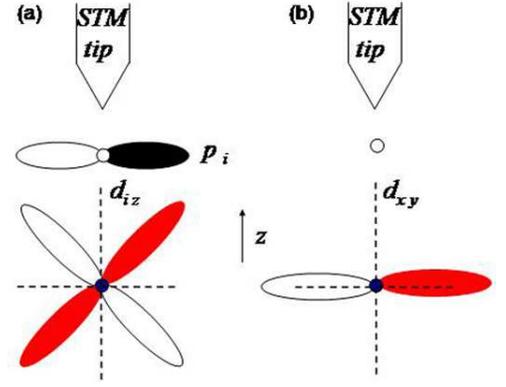,clip=1,width=0.75\linewidth,angle=0}
\caption{\label{fig:tunneling} (Color online) Schematic illustration of the wave function overlap related to the tunneling matrix element for STM tip.
The tunneling of electrons from the STM tip to the $d$-orbitals of the Ru atoms must go through the oxygen atoms (white dots).
(a) For $d_{iz}$ orbitals ($i=x,y$), the tunneling matrix element is large with the help of the $p_i$ orbital of the oxygen atoms.
(b) For $d_{xy}$, all p-orbitals of the oxygen atoms have zero wave function overlaps with it, leading to much weaker
tunneling matrix element compared to $d_{yz,xz}$ orbital.}
\end{figure}

It can be seen that  Fig. \ref{fig:qpi}(c) alone captures the main features of the experimental results of Ref. \ocite{lee2009qpi327},
suggesting that the contribution from the 2d $d_{xy}$ band is essentially invisible in SI-STM experiment.
The missing of $d_{xy}$ band scatterings in the experiment can be explained by appealing to the aforementioned orbital-dependence
of the STM tunneling matrix elements.  Because the surface of the material is usually cleaved such that the outermost layer is the oxygen layer, 
there is an oxygen atom lying above each uppermost Ru atom.  As a result, the tunneling matrix element will be mostly determined by the
wavefunction overlaps between the $p$-orbitals of the oxygen atom and the $d$-orbitals of the Ru atom.
As illustrated in Fig. \ref{fig:tunneling}, the wavefunction overlaps of the $d_{yz}$ ($d_{xz}$) orbital with the $p_y$ ($p_x$) are large
while none of the $p$-orbitals has finite overlaps with $d_{xy}$ orbital, leading to $C^{xy} \ll C^{yz}=C^{xz}$.
Moreover, the tunneling matrix elements also depend on the in-plane momentum $\kvec$.
It has been shown theoretically that the tunneling matrix elements 
have important effects in the tunneling spectra \cite{tersoff1983,
wu2002,zhang2008,chuang2010}.
These matrix elements
are significantly suppressed at large in-plane momentum\cite{tersoff1983},
and recent STM experiments on graphene\cite{zhang2008} and iron-pnictide superconductors\cite{chuang2010} have demonstrated this suppression.
Since the $\gamma_{1,2}$ sheets are located around momenta much larger than those of $\alpha_{1,2}$, their contributions could be further
suppressed by this effect.

Based on the above discussion, we will henceforth set $C^{yz}=C^{xz}=1$ and $C^{xy}=0$.

\subsection{QPI imaging at energy below the Fermi energy}

\begin{figure}
\centering\epsfig{file=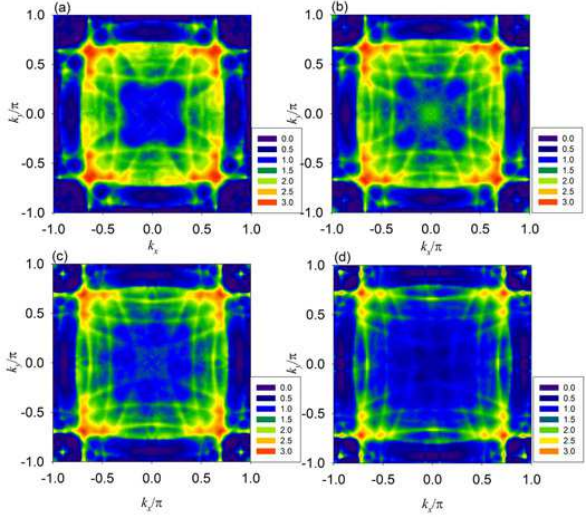,clip=1,width=0.9\linewidth,angle=0}
\caption{\label{fig:qpi-dispersion} (Color online) QPI imaging at (a) $E=0$, (b) $E=-0.03$, (c) $E=-0.06$, and (d) $E=-0.1$.
The main features are similar because the Fermi surfaces of $\alpha_{1,2}$ are relatively insensitive to $E$ throughout this energy range.}
\end{figure}

Since the experiments were done at negati ve sample bias voltage \cite{lee2009qpi327}, we compute the QPI imaging
for several negative values of $E$.  Fig. \ref{fig:qpi-dispersion} present the QPI imaging for $E=0, -0.03, -0.06, -0.1$, and the
main features of the stripe-like patterns remain unchanged. 
This is also consistent with the experiments showing that the QPI imaging are similar for sample bias voltage down to $E=-12\,$meV, 
and the reason is that the Fermi surfaces of $\alpha_{1,2}$ do not change very much throughout this range of energy.

\subsection{QPI imaging for impurities at different layers}

\begin{figure}
\centering\epsfig{file=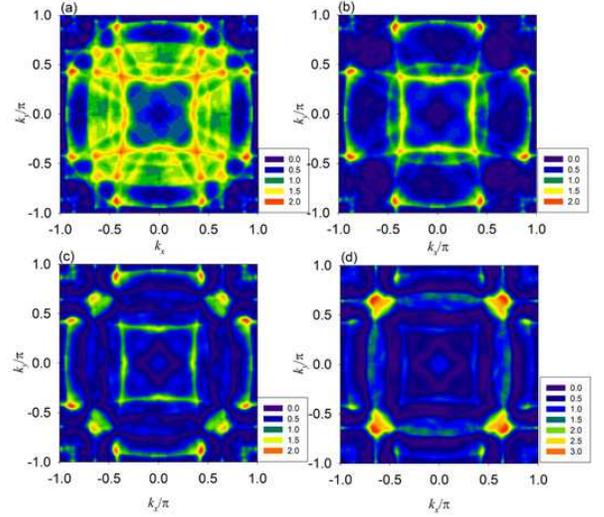,clip=1,width=0.9\linewidth,angle=0}
\caption{\label{fig:qpi-real} (Color online) QPI imaging evaluated from Eq. \ref{realdidv} for (a) $x=0.25$, (b) $x=0.5$, (c) $x=0.75$, and (d) $x=1$.
The QPI imaging in (a) fits the experimental result the best.}
\end{figure}

The above calculations were all performed assuming that the scattering impurity is located on the top layer only.
However, QPI from impurity scattering in the second layer may also be detectable in experiments.
Since the measurements of the conductance $dI/dV$ are more likely an average of both cases, it is reasonable to expect
\be
\frac{dI}{dV}(\vec{r},E)\propto (1-x)\rho_{\ssr{TOP}}(\vec{r},E) + x \rho_{\ssr{BOTTOM}}(\vec{r},E)\ ,
\label{realdidv}
\ee
where $\rho_{\ssr{TOP}}(\vec{r},E)$ is the LDOS of quasi-1d bands with impurity on the top layer, 
and $\rho_{\ssr{BOTTOM}}(\vec{r},E)$ is that with impurity on the bottom layer.
We can then obtain the QPI imaging by performing a Fourier transformation on Eq. \ref{realdidv} as a function of $x$.  The results
are presented in Fig. \ref{fig:qpi-real} for $x=0.25,0.5,0.75,1$. 
We find that $x=0.25$ best reproduces the experimental data of Ref. \ocite{lee2009qpi327}.

\section{Implication of orbital ordering from QPI imaging}
\begin{figure}
\centering\epsfig{file=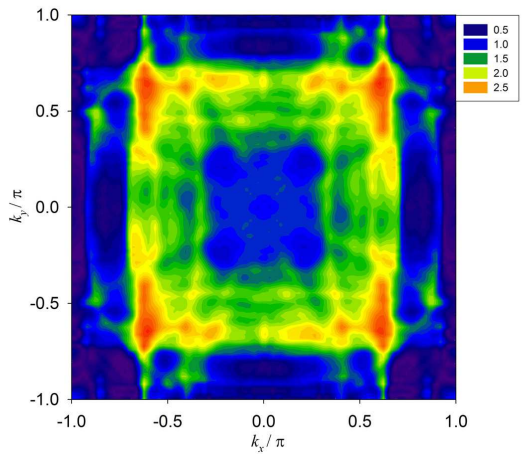,clip=1,width=0.8\linewidth,angle=0}
\caption{\label{fig:nematic} (Color online) QPI imaging with nematic order. $N/t_1=0.1$ and $\mu_B B/t_1=0.06$ is chosen.
The breaking of the $C_4$ symmetry to $C_2$ symmetry is clearly seen.}
\end{figure}

Two of us\cite{lee2009} have proposed that the nematic order observed in this material results from an orbital ordering in the quasi-1d bands
enhanced by the orbital hybridizations.
The charge and spin nematic order parameters $n_c$,$n_{sp}$ can be expressed as:
\bea
n_c=\frac{1}{2}\big(\langle n_{xz}\rangle-\langle n_{yz}\rangle\big)\,\,\,,\,\,\,n_c=\big(\langle S^z_{xz}\rangle - \langle S^z_{yz}\rangle \big).
\eea
The mechanism of the nematic order under the magnetic field is that the majority spin band is pushed closer to the van Hove
singularity, which triggers the nematic distortion in the majority spin Fermi surfaces.
The mean field theory\cite{lee2009} with a microscopic model of quasi-1d bands also reproduced this feature, leading to $n_c=n_{sp}$.
To calculate the QPI imaging with a nematic order, we introduce two new terms into the Hamiltonian:
\bea
H_{\rm nematic}&=&N\sum_{\rvec,a} \big(d^{yz\, \dagger}_{\uparrow,a}(\rvec) \, d^{yz}_{\uparrow,a}(\rvec) - 
d^{xz\, \dagger}_{\uparrow,a}(\rvec) \, d^{xz}_{\uparrow,a}(\rvec)\big)\nn\\
H_{\rm Zeeman}&=& -\mu_{\rm B} B\sum_{\rvec,a,\alpha} \big(d^{\alpha\, \dagger}_{\uparrow,a}(\rvec) \, d^{\alpha}_{\uparrow,a}(\rvec)
- d^{\alpha\, \dagger}_{\downarrow,a}(\rvec) \, d^{\alpha}_{\downarrow,a}(\rvec)\big)\ ,\nn
\eea
where $N=n_{\rm c}+n_{\rm sp}$ measures the strengths of the nematic distortion in the majority spin Fermi surfaces.
Fig. \ref{fig:nematic} shows the QPI imaging at $E=0$ with $N/t_1=0.1$ and $\mu_{\rm B} B/t_1=0.06$.
As expected, a stripe-like pattern breaking the $C_4$ symmetry down to $C_2$ symmetry is observed.

We propose that this result could be used to resolve the controversy on which band is responsible for the nematic order.
If the nematic order occurs mostly in the $d_{xy}$ band and the quasi-1d bands do not exhibit orbital ordering,
the QPI imaging from the experiments should have a $C_4$ symmetry even within the range of the nematic order
because the SI-STM is not sensitive to the $d_{xy}$ band.
Conversely, if the orbital ordering in the quasi-1d bands is responsible for the nematic phase, the SI-STM will see
the imaging with only $C_2$ symmetry, as shown in Fig. \ref{fig:nematic}.

\section{Conclusions}
\label{sect:conclusions}

In this paper, we have constructed a bilayer tight-binding model with 
three $t_{2g}$ orbitals for the Sr$_3$Ru$_2$O$_7$, with careful attention 
paid to details of the lattice structure.  
We found that the rotations of the in-plane octahedra oxygens induce 
new hoppings between quasi-1d $d_{yz}$ and $d_{xz}$ bands with staggered 
signs in the hopping integrals, which in turn lead to a unit cell 
doubling consistent with what is observed in both ARPES 
experiment\cite{tamai2008} and LDA calculations\cite{tamai2008,singh2001}.
This mechanism for unit cell doubling is distinct from that in the 
model used by Puetter {\it et al.\/}\cite{puetter2010},
in which a staggered on-site potential is introduced to distinguish 
the sublattices.  
Furthermore, we have also computed the quasi-particle interferences 
in the spectroscopic imaging STM based on a multiband $T$-matrix 
approach within this tight-binding model.
Due to the effects of tunneling matrix elements, we find that the the QPI 
imaging measured by Lee {\it et. al.}\cite{lee2009qpi327} are dominated 
by the scatterings in the quasi-1d $d_{xz}$ and $d_{yz}$ bands, and the 
contribution from the 2-D $d_{xy}$ band is largely suppressed.
We have further considered the possibility of impurities residing on 
either top or bottom layers, and a linear combination of these two cases 
leads to the best fit with the experiments.

We have also calculated the QPI imaging for the system with a orbital ordering in the quasi-1d bands in a magnetic field, and 
we propose that this could be a realistic way to distinguish which band is responsible for the nematic order.
We predict that if the $d_{xy}$ band is the dominant band for the nematic phase and no orbital ordering in quasi-1d bands is present, 
the QPI imaging will still preserve the $C_4$ symmetry even within the nematic phase because the SI-STM could not detect the change
in the $d_{xy}$ band.
On the other hand, if the orbital ordering in quasi-1d bands is responsible, a breaking of the $C_4$ symmetry down to $C_2$
should be observed in the QPI imaging as the system enters the nematic phase. 

One remarkable aspect in our tight-binding model is the introduction of $V_{\rm bias}$, the difference in on-site potential for the top and bottom layers. 
It has been shown here that the crossings of the Fermi surfaces with different 'layer parities' can not be avoided without a $V_{\rm bias}$ term. 
In order to reproduce the Fermi surface sheets mapped out from the ARPES experiments, especially for $\alpha_2$, a non-zero $V_{\rm bias}$ is essential.  Physically since the ARPES still measures mostly the electronic properties near the surface, it is reasonable to expect that 
the surface work function could produce a sizable $V_{\rm bias}$ to be seen in the ARPES.
Furthermore, the fact that STM, another surface sensitive probe, detected only the $\alpha_2$ band scatterings also supports the 
existence of a non-zero $V_{\rm bias}$. 
On the other hand, $V_{\rm bias}$ vanishes inside the bulk, and thus 
the bulk Fermi surfaces would have different shapes and volumes from 
those obtained by ARPES \cite{tamai2008}.
This issue is important when comparing the Fermi surfaces measured in ARPES
with those measured in quantum oscillations experiments, since the former is a surface measurement while the latter is a bulk one.


\section{Acknowledgement}
We thank A. Mackenzie and J. C. Davis for helpful discussions.
C. W. and W. C. L. are supported by ARO-W911NF0810291 and Sloan 
Research Foundation.



\end{document}